\documentstyle[12pt,preprint,aps]{revtex}

\newcommand{\eq}[2]{\begin{equation} #1 \label{eq:#2} \end{equation}}
\newcommand{\req}[1]{(\ref{eq:#1})}
\newcommand{\f}{\frac}
\newcommand{\ph}{\phi}
\newcommand{\dc}{\delta_{4}}
\newcommand{\fjp}{\frac{1}{2}}

\begin{document}

\draft

\title{
 THE PHASE DIAGRAM FOR THE SINE-GORDON MODEL WITH TWO
 UMKLAPP TERMS}

\author{ A. Bjeli\v{s} and  M. Latkovi\'{c} }

\address{
 Department of Theoretical Physics, Faculty of Sciences\\
 University of Zagreb,  P.O.B. 162, 41001 Zagreb, Croatia\\ }

\maketitle
\begin{abstract}
We study the Landau free energy for a uniaxial ordering, taking into account
two Umklapp terms of comparable strengths (those of the third and fourth
order). Exploring the analogy with the well-known nonintegrable classical
mechanical problem of two mixed nonlinear resonances, we complete the
previous studies of the corresponding phase portrait  by calculating
numerically periodic solutions, including those far from the separatrices.
It is shown that in the physical range of parameters only periodic
configurations are absolutely stable. We determine for the first time the
complete thermodynamic phase diagram and show that, in contrast to some
earlier claims, the wave number of the ordering does not pass through the
devil's staircase,  but through a finite number of steps which decreases
as the amplitudes of the Umklapp terms increase.
\end{abstract}


The sine-Gordon model~\cite{mcm,bul} which is widely used in the studies of
the  uniaxial incommensurate-commensurate (IC) transitions~\cite{cum},  is
based on a few intricate approximations. By one of them only one, presumably
dominant, phase dependent Umklapp term is retained in the Landau expansion
of the free energy. In this Letter we critically examine this approximation
by including two Umklapp terms which favor different commensurabilities, {\em
i.e.} those of the third and fourth order.
Like in other extensions of the sine-Gordon model~\cite{aub,bak,joo,fra,err},
one encounters a nonintegrable  problem, well-known in the modern classical
mechanics~\cite{chi,esc}. As a consequence, the thermodynamic phase diagram
for the IC transition qualitatively differs from that of the simple
sine-Gordon model.

It is well-known that continuous nonintegrable models~\cite{fra,err} and
discrete models~\cite{aub,bak,joo} for the IC transition lead to similar
descriptions of uniaxial orderings. The analyses were however usually made
for limited ranges of values of physical parameters which enter into models.
Also, the continuous models were examined only in the parts of phase space
close to separatrices, characterized by dilute soliton lattices.
With such restrictions it was not possible to determine complete thermodynamic
phase diagrams which are the basis for any discussion of particular real
systems.
In this Letter we combine two numerical methods, with the aim to investigate
the entire phase space and to find the regions of absolute stability for
particular solutions in the plane spanned by two control parameters (out of
three present in the model). It will be shown that, due to the gradual
disappearance of quasiperiodic solutions and the formation of chaotic layers,
the phase diagram is (almost) entirely covered by periodic configurations.

The model is defined by the free energy functional
\eq{ F=\int dx[\fjp (\f{d\ph}{dx})^{2}+B\cos{(3\ph+3(\f{\pi}{6}-\dc)x)}+
                C\cos{(4\ph-4\dc x)}] }{1}
with  $\phi$  being the phase of the order parameter.  $\delta_4 \equiv  Q -
Q_4$,
where $Q$ is the wave number of ordering in the absence of Umklapp terms
and  $Q_4 = 2\pi/4$ is the commensurate wave number of the fourth order.
(We put everywhere the lattice constant equal to unity.)
By assumption $Q$ is placed somewhere between $Q_4$ and the commensurate
wave number of  the third order $Q_3 = 2\pi/3$. In that case the Umklapp
terms of third and fourth order which are included into the Landau
expansion~\req{1} might be of comparable strengths. {\em E. g.} the Umklapp
term of $m$-th order for charge and spin density waves is proportional to
the $m$-th power of the amplitude of the order parameter~\cite {lee}.
Thus, at temperatures not far below the critical temperature $T_c$ of the
incommensurate ordering, the third order Umklapp term should not be omitted
due to the dominance of the coupling constant $B$ with respect to $C$. On the
other hand, for small values of $\dc$  ({\em e. g.} for $\dc \ll \pi/6-\dc$)
the closeness of the wave number to the commensurate point $Q_4$ favors the
fourth order Umklapp term.
The similar arguments can be invoked for other types of uniaxial orderings.
Note that the example~\req{1} is the most interesting one since it includes
two Umklapp terms which are of the lowest possible orders in the amplitude
of the order parameter. (The model for the commensurate lock-in at $Q = \pi$
is of essentially different type~\cite {mic,dan}).

The competition between two periodicities in the expansion~\req{1} takes place
in the nonlinear terms which are formed on the purely symmetry grounds. It is
present in {\em both} the $\phi$ dependence and
the "nonautonomous" $x$-dependence of these terms. We recall that the
starting functional  of Fradkin et al~\cite{fra}, which is appropriate for
the approximate treatment of the ordering in the presence of mobile
defects~\cite{err}, contains only the competition of the latter type.

The extremalization of the functional~\req{1} leads to the Euler-Lagrange
(EL) equation
\eq{ \ph''+3B\sin{[3\ph+3(\pi/6-\dc) x]}+4C\sin{(4\ph-4\dc x)}=0 .}{2}
Among all  solutions  of this equation we are looking for that which is the
absolute minimum of the mean free energy density in the eq.~\req{1}~\cite{co1}.
To this end we develop a procedure which is complementary to
those usually performed in the classical mechanical analyses of the
phase portrait of  eq.~\req{2} (see {\em e. g.} refs.~\cite {chi,esc}).
Standard methods of direct integration of eq.~\req{2}, which were widely used
in the calculations of quasiperiodic and chaotic orbits,
cannot be extended to the isolated periodic solutions,  which are orbitally
unstable in the most interesting cases of phase portraits with finite chaotic
layers.  However, it appears that just these solutions have the main role in
the thermodynamic phase diagram. We therefore calculate them separately,
by treating eq.~\req{2} as the boundary value problem.
Combining these two methods we can follow all relevant trajectories in the
phase portrait and extract the trajectory for which the functional~\req{1}
has the absolute minimum.

The functional~\req{1} and the EL equation~\req{2} do not possess the
continuous translational symmetry in $x$, but are invariant under
the combinations of discrete translations in {\em both} $x$
and $\phi$~\cite{lat}.
All solutions of eq.~\req{2} can be thus grouped into degenerate subsets
of solutions connected by these translations. In the further
considerations of the periodic solutions we enumerate these subsets
and represent each subset by one solution.

It is easy to determine possible periods of the periodic solutions, defined by
\eq{ \ph(x+P)=\ph(x)+\ph_{P}. }{3}
Inserting eq.~\req{3} into the EL equation~\req{2}, one gets the  necessary
conditions
\eq{P = 4k + 3l,  \hspace*{1cm} \phi_P = \delta_4 P - l\pi/2,}{4}
where $k$ and $l$ are integers.
The values of $P$ are  $\phi_P$ are thus discretized  (in contrast to the
simple sine-Gordon case) and denote allowed intervals of variables $x$
and $\phi$ in the boundary value problem. Note that the solutions with
finite values of $\phi_P$ are rotational ({\em i.e.} have finite winding
numbers).

The boundary conditions for the periodic solutions with the values of  $P$
and  $\phi_P$ given by eq.~\req{4} are defined
by $\phi_{kl} (x=x_i)$ , $\phi_{kl} (x=x_f=x_i + P) = \phi_{kl} (x=x_i) +
\phi_P$ and  $\phi_{kl}'(x=x_i) = \phi_{kl} '(x=x_f)$.
Using two different numerical methods~\cite{num}
we undertake a systematic search of such conditions for periods $P$ up
to $\approx 100$. Both numerical procedures give at most two solutions with
the same values of $k$ and $l$, defined by two isolated points in the
two-dimensional plane of boundary
conditions  $[\phi_{kl} (x=x_i),  \phi _{kl}'(x=x_i)]$~\cite{co2}.
We noticed~\cite{lat} that these points sometimes have jumps and
discontinuities as the functions of the parameters $B$ and $C$. In particular,
a given periodic solution may cease to exist as the nonlinearity
strengthens, the general tendency being that the
solutions with smaller periodicities are more robust as $B$ and/or $C$
increase.

When both exist, two solutions with same values of $P$ and $\phi_P$ have
different values of the free energy~\req{1}. Both have odd symmetry
[the only possible for rotational solutions], but
with different positions of inversion points [$x_0,  \phi(x_0)$]. It easily
follows from the differential equation~\req{2} that $x_0$  necesarily has only
integer and half-integer values,
while $\phi(x_0)=x_0\delta_4-m\pi/4$, \ \ \mbox{$m = 0, \pm1, \pm2, ... $.}
This is indeed the case with all our numerical solutions.
With the odd symmetry taken into account, the numerical procedure
is further simplified, since it remains to determine only one continuous
boundary condition [$\phi _{kl}'(x=x_0)=\phi _{kl}'(x=x_0+P)$] and two
discrete numbers [$x_0$ and $\phi(x_0)$] per each solution.
This analysis~\cite{lat} shows that the particular
values of $x_0$ and $\phi(x_0)$ for a given solution are related to the
evenness and oddness of the corresponding integers $k$ and $l$.
Further below we follow only periodic solutions with the
lower free energy~\req{1}, and characterize each such solution by a pair of
integers ($k, l$).

The structure factor for a given periodic solution ($k, l$) has equidistant
satellites, whose positions with respect to the the wave number $Q$
are given by
\eq{q_n\equiv\f{\phi_P}{P}+\f{2n\pi}{P}=\dc+\pi\f{2n-l/2}{4k+3l}, \hspace*{1cm}
n = 0, \pm 1,  \pm 2,  ...} {5}
The position of the  fundamental satellite $q_0$  is determined by the total
slope of the function $\phi_{kl}(x)$, while the higher harmonics $q_n, n\neq 0$
appear due to its periodic modulation.

In the next step we determine mean free energies of particular solutions. For
illustration, we show in Figs.1a-c mean free energies of periodic,
quasi-periodic and chaotic solutions with one fixed [$\phi (x_i = 3/2) = 0$]
and one variable  [$\phi_0^{'}\equiv \phi '(x_i = 3/2)$]
initial condition, choosing $\dc =\pi/12$ and three representative values
of  $B$ and $C$.
For all periodic solutions shown in these figures $x _i= 3/2$ coincides with
the point of odd symmetry $x_0$. Since these figures remain qualitatively
same for other choices of the "initial" point $x_i$, they enable some general
conclusions on the relationship between free energies of various types of
solutions.

The situation in which the parts of the phase space with the rotational flow
are still free of chaos is presented in Fig.1a. It is evident
that there is a periodic solution which has lower mean free energy
than all other calculated periodic and quasiperiodic solutions.
The same numerical conclusion is obtained for other finite values
of $B$ and $C$.
It should be however mentioned that the comparison of mean free energies of
periodic and quasiperiodic configurations becomes more and more subtle
as $B$ and/or $C$ decrease, so that in this limit one cannot rigorously exclude
the absolute thermodynamic stability of some quasiperiodic
solution~\cite{bak,joo,fra}.

We turn from this  intricate and numerically unpleasant limit to the
intermediate values of $B$ and/or $C$ for which both, quasiperiodic and
chaotic, layers are present in the rotational part of the phase
space [Fig.1b], and finally to the regime~\cite{chi}
\eq{\sqrt{B} + \sqrt{C} > \pi/12} {6}
in which there is no quasiperiodic solution and this part is covered by a
single chaotic layer (Fig.1c). [Note that the wings on  the left and right
sides in Figs.1.a-c come from the thermodynamically uninteresting oscillating
solutions of eq.~\req{2}.]
Since the numerical integration of eq.~\req{2} cannot extract a single chaotic
solution (due to its  orbital instability), the {\em plateaux} in Figs.1b,c
strictly represent mean free energies within a given chaotic layer, {\em
i.e.} each {\em plateau} represents one isolated chaotic layer in the phase
space. Still, it is plausibly expected~\cite{fra} that each particular chaotic
solution fully diffuses inside "its" layer, so that its mean free energy
coincides with one of the {\em plateaux} in Fig.1. Periodic solutions are
immersed into these chaotic layers as isolated points. Their free energies
are discontinuous (needle-like) minima well below all chaotic {\em
plateaux} [see Figs.1b (inset) and 1c].
As it  was already noticed, the number of surviving periodic solutions
decreases with the increase of "chaoticity" of the phase space, measured by
the left-hand side of the inequality~\req{6}.

The numerical results illustrated by Figs.1a-c are the basis for the
construction of the thermodynamic phase diagram in the three-dimensional
space of parameters $B, C$ and $\delta_4$. One possible visualization is
presented in Figs.2  and 3 in which we fix  $B$,  vary $C$
and $\dc$~\cite{co3}, and follow the fundamental wave
number $q_0$ [eq.~\req{5}] and the free energy of the configuration with
the lowest value of the functional~\req{1}. For $C$ large enough [{\em e. g.}
such that the inequality~\req{6} is satisfied],  the system passes through
small number of periodic configurations as $\dc$ varies. Other existing
periodic solutions have free energies which are larger and increase as the
period $P$ increases~\cite{lat}.  Since it does not seem probable that the
further solutions, which have even higher periods
and were not calculated numerically, would not follow this tendency, we
conclude that there are no other solutions which would fill the phase
diagram between the neighboring {\em plateaux} shown in the Fig.2.

As $C$ decreases the number of periodic configurations increases through
a series of bifurcations, while the dependences
of the free energy and the wave number on  $\dc$, although discontinuous,
approach that of the simple sine-Gordon limit $C = 0$.
The number of absolutely stable periodic configurations still remains finite
(and smaller than the number of numerically calculated periodic
solutions), even for $C \approx 0.001 - 0.002$ when there are no chaotic
layers in the rotational part of phase space (except those in the regions of
separatrices) (Fig.1a). In other words, our numerical results suggest that
the staircase from Fig.2 remains "harmless"~\cite{vil} ({\em i.e.} not
fractal) as long as $C$ is finite (or at least not too small).
The harmless staircase with first order transition between neighboring
periodic solutions is expected to be closely connected with the participation
of dense soliton lattices in our phase diagram. Then it is inappropriate
to use the picture of well separated solitons with a repulsive effective
interaction, which was usually the basis for the conclusions that the wave
number $q_0$ may pass through the complete or incomplete devil's
staircase ~\cite{joo,fra}.

In conclusion, we stress that the periodic solutions are of the central
importance for the thermodynamics of  systems modelled by the free
energy~\req{1}. In the physical range of parameters ($B$
and $C$ finite), only these solutions ({\em i.e.} the commensurate
configurations)  have the property of absolute thermodynamic stability.
In the phase space they are isolated and immersed in a mostly chaotic
environment, whose mean free energies represent characteristic
energetic barriers which the system has to overcome in order to replace one
periodicity by another. Such barriers might be intrinsic sources~\cite{bje}
of memory effects and thermal hystereses which are frequently observed in
uniaxial systems with a variable wave number of ordering~\cite{cum}.
We recall that the thermodynamic significance of periodic and chaotic
solutions is just opposite to that of the classical mechanical counterpart of
the model~\req{1}. Mechanical periodic trajectories cannot be realized due to
their orbital instability, so that the rotational motion of a strongly
nonintegrable classical mechanical system is entirely chaotic.

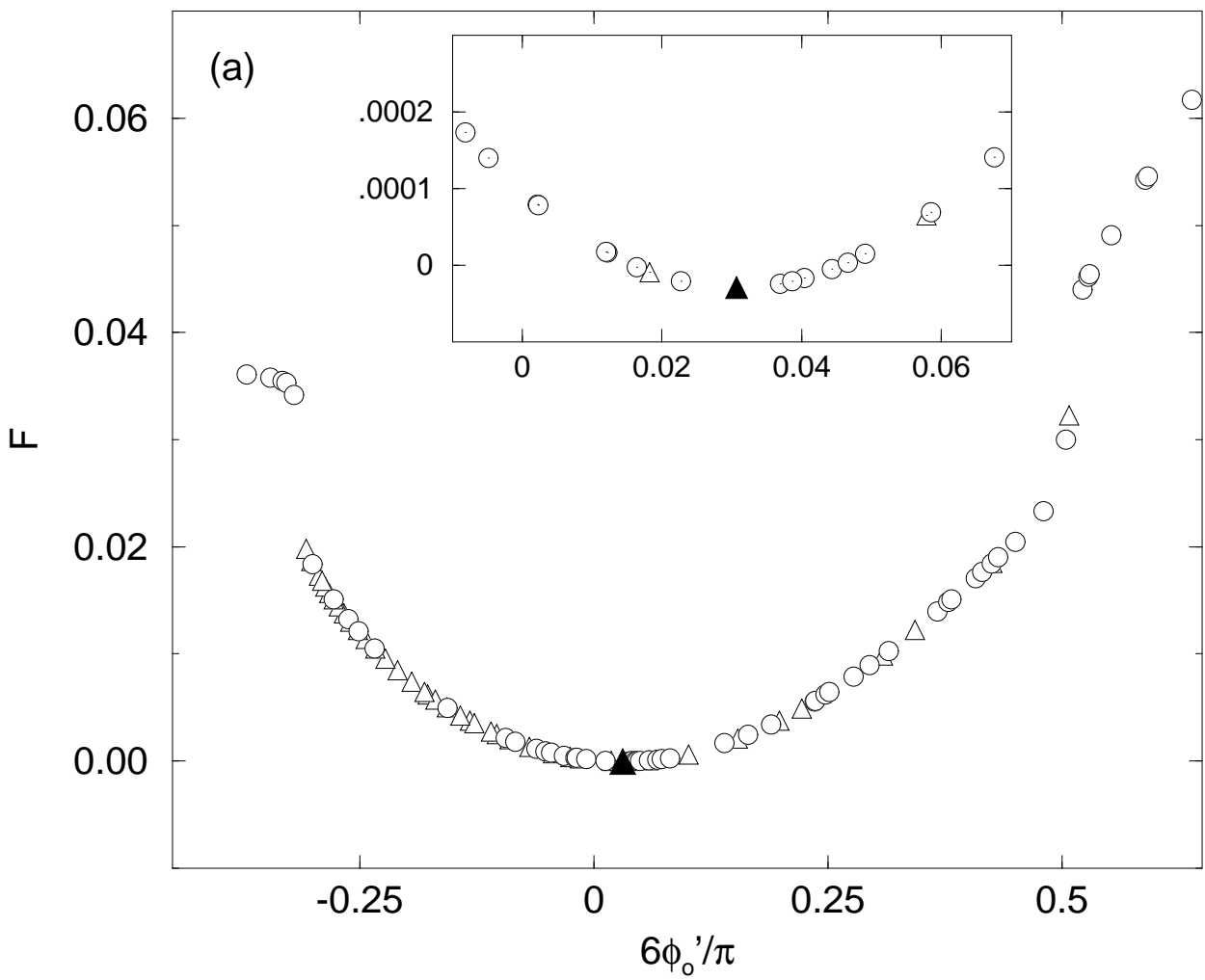
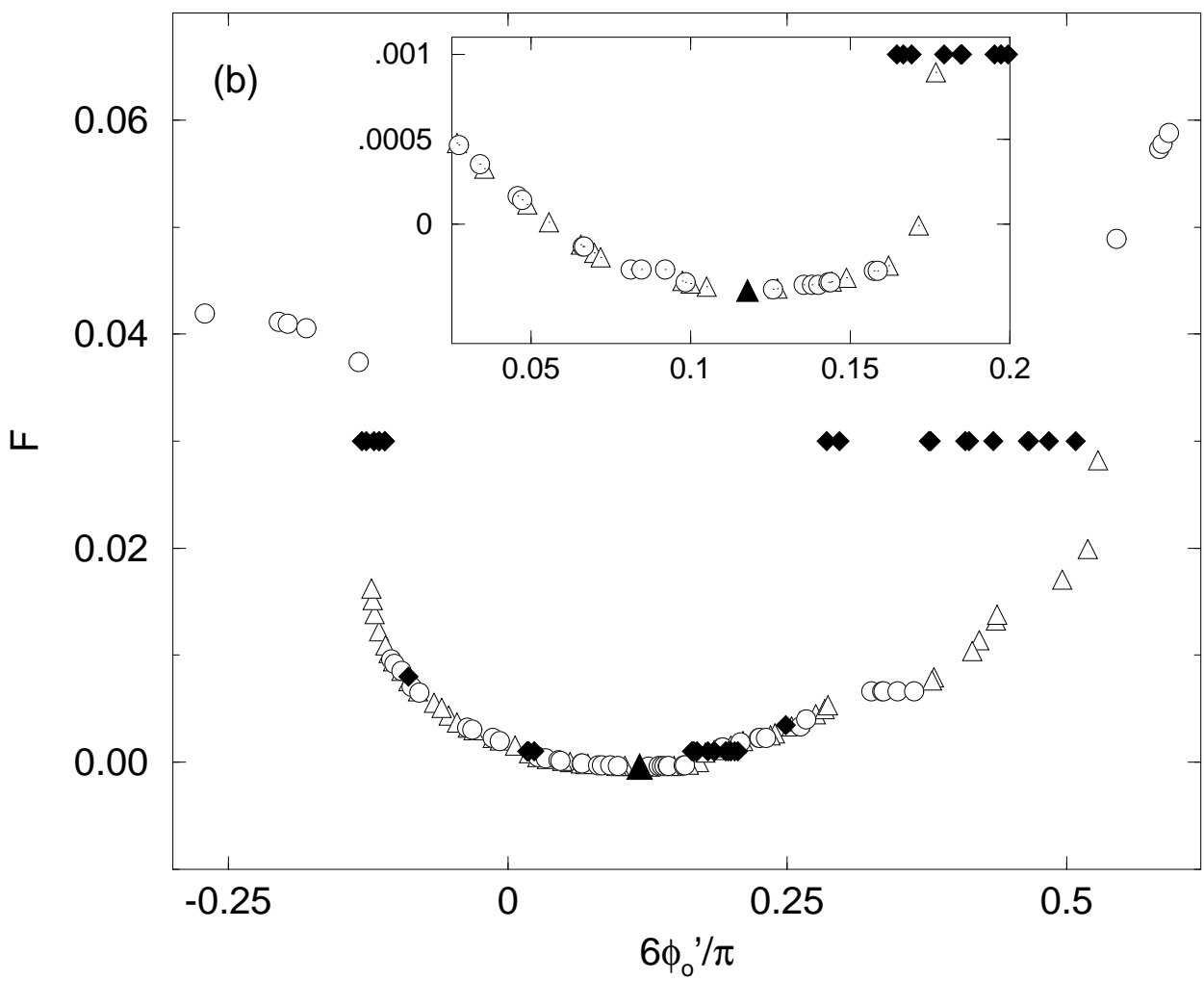
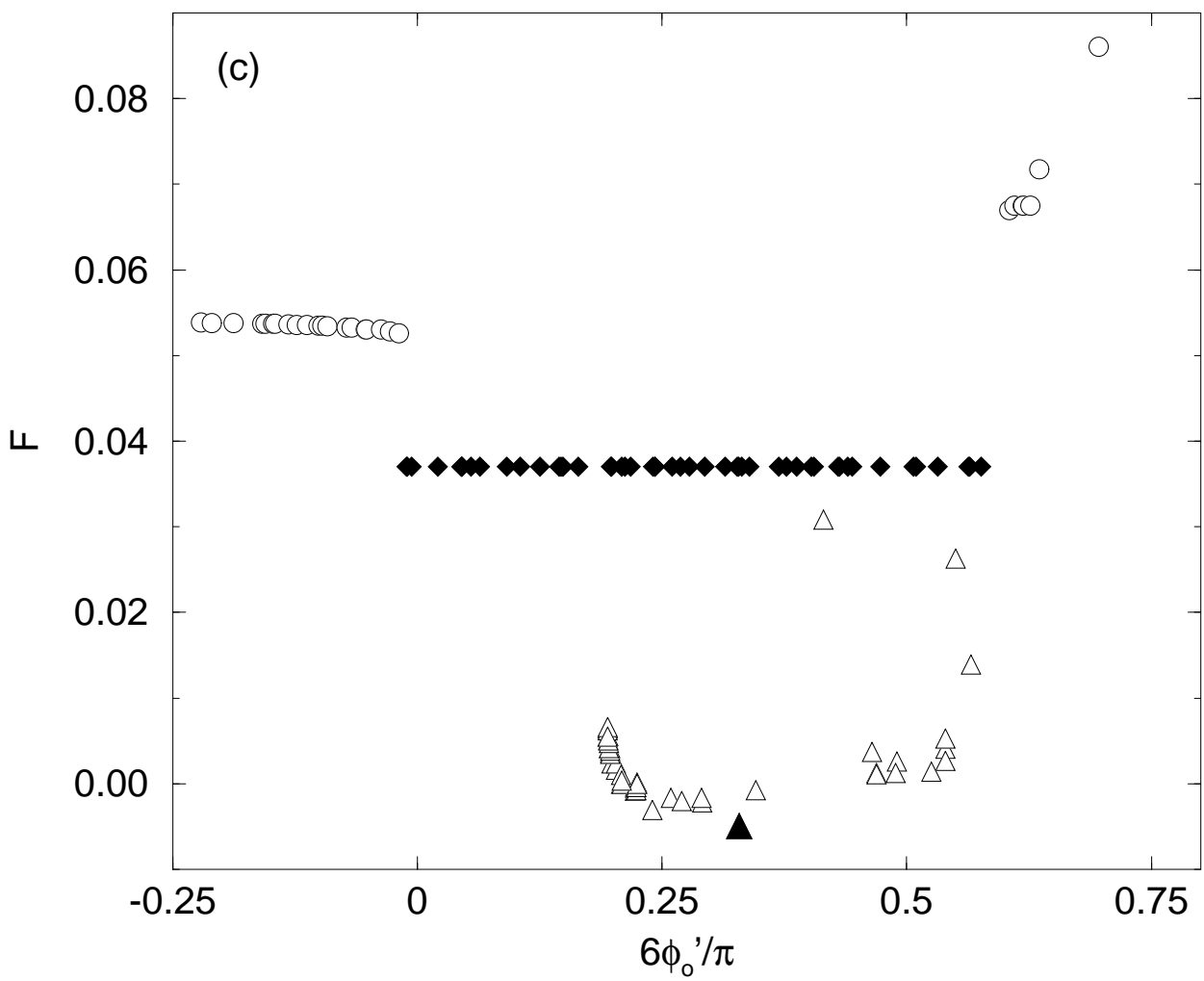
\begin{figure}
\caption{
Mean free energy {\em vs} $\phi_0^{' }$ for $\delta_4 = \pi/12$, $B = 0.002,
C = 0.002$ [a], $B = 0.008, C = 0.006$ [b] and  $B = 0.02,  C = 0.02$ [c] of
periodic ($\triangle$), quasiperiodic ($\bigcirc$) and chaotic
($solid \Diamond$) solutions. The ($k, l$) indices for the periodic solution
with the lowest free energy ($ solid \triangle$) are $(3, 4)$ in the
figures [a] and [b], and $(1, 1)$ in the figure [c]. The insets in
figures [a] and [b] are enlarged neighborhoods of the free energy minima.}
\label{fig1}
\end{figure}

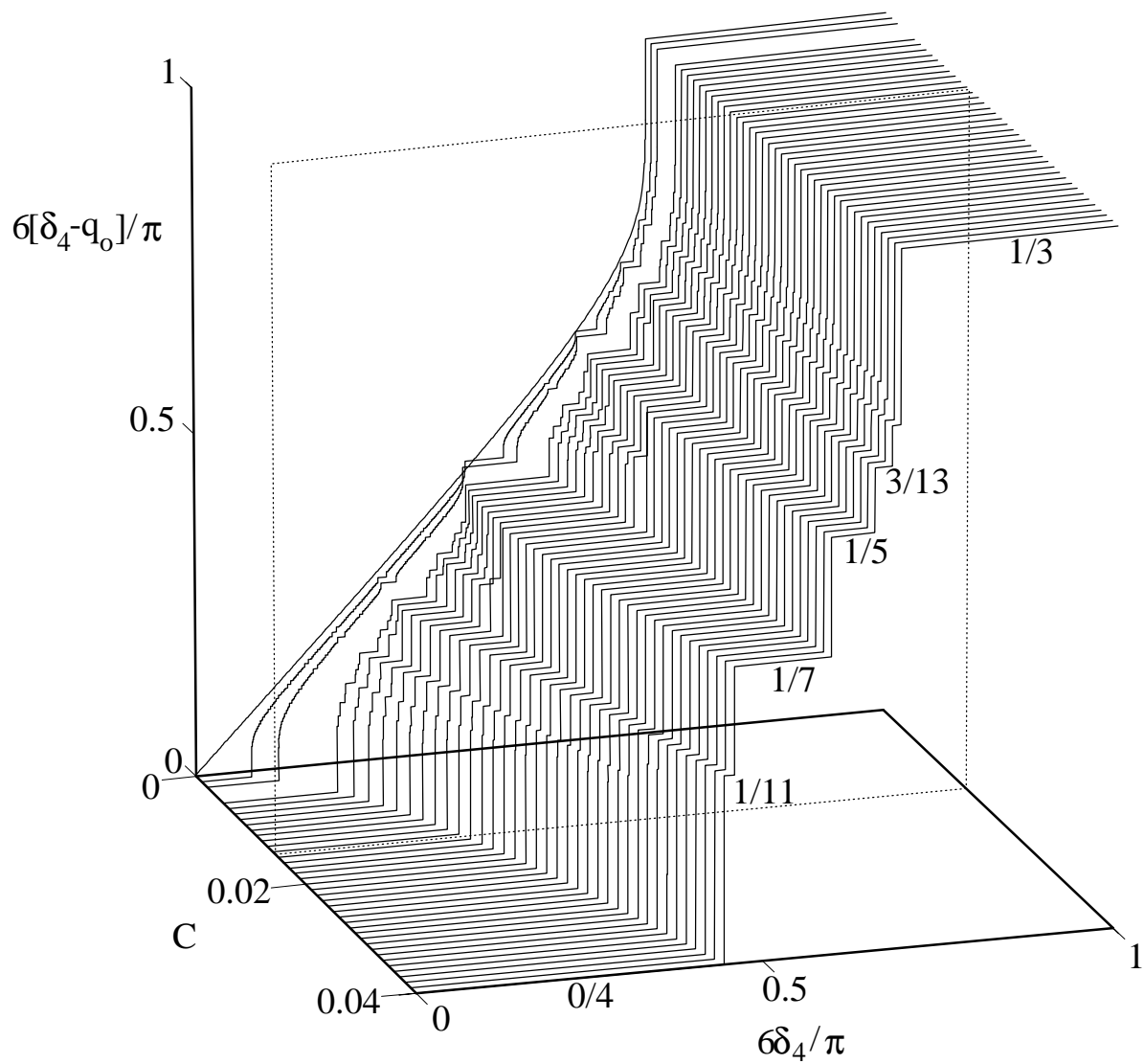
\begin{figure}
\caption{
The fundamental wave number $q_0$ {\em vs} $C$ and $\delta_4$  for $B = 0.02$.
The fractions at particular steps denote the
ratios $2(\dc-q_0)/\pi = l/(4k + 3l)$. In front of the dotted
cross-section the rotational part of the phase space does not contain
quasiperiodic solutions but only a single chaotic layer and isolated
periodic solutions [see eq.~\protect{\req{6}}].}
\label{fig2}
\end{figure}

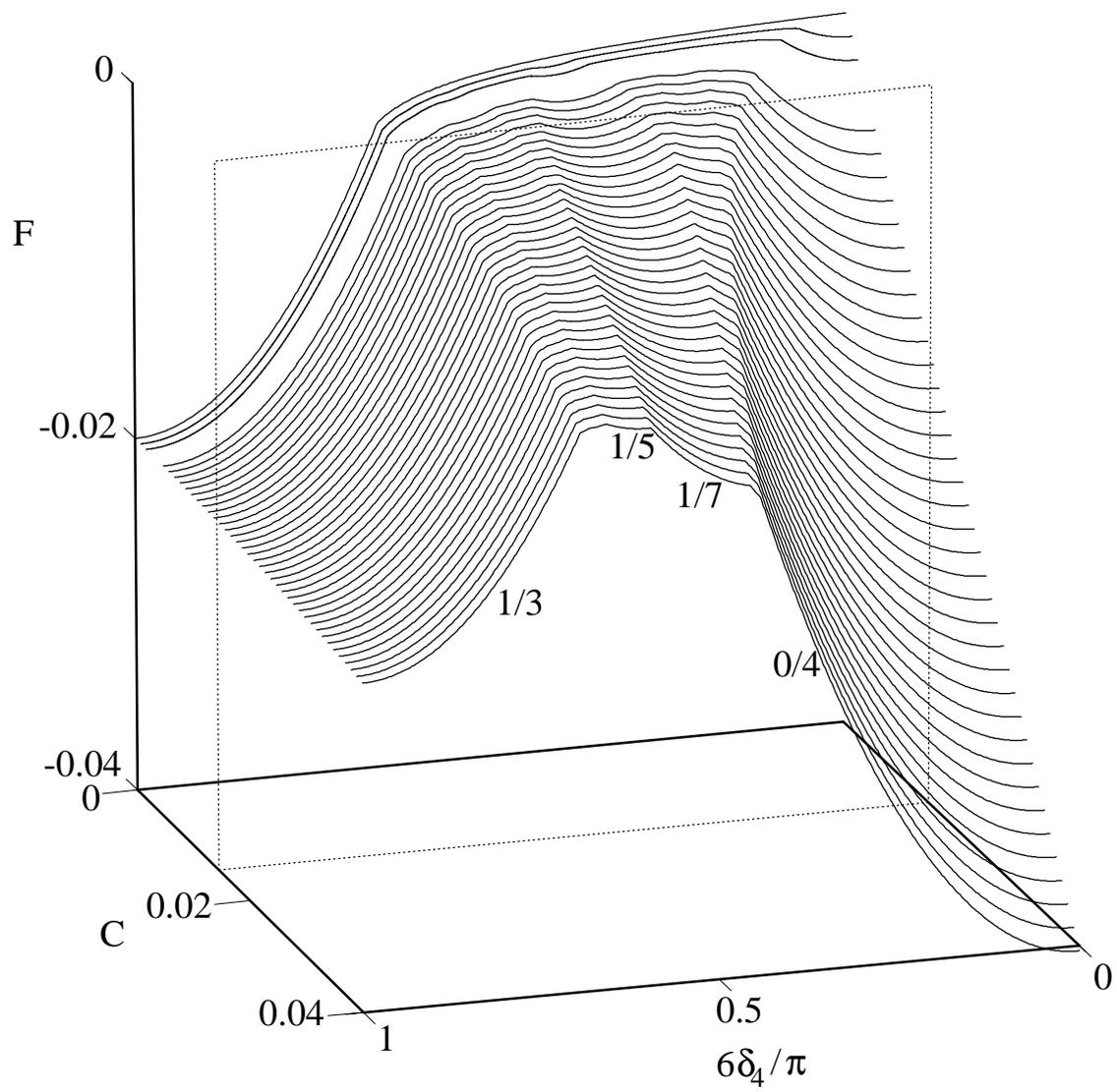
\begin{figure}
\caption{
Free energy of the absolutely stable periodic configurations {\em vs} $C$
and $\delta_4$  for $B = 0.02$. The fractions and the dotted cross-section
have the same meaning as in Fig.2. Note the reversed
scales at the $\delta_4$ axes in Figs.2 and 3.}
\label{fig3}
\end{figure}


\begin{references}
\bibitem{co2} In principle the choice ($mk, ml$) where m is some integer may
either just reproduce the same periodic solution or represent a new one
with $m$-times larger values of $P$ and  $\phi_P$. Only the former possibility
is realized  numerically, so  that all distinct periodic solutions are
characterized by pairs  of integers ($k, l$) which do not have a common
integer factor.
\bibitem{co3} The phase diagram in the ($B, C$) plane with $\delta_4$ fixed
which could be more useful for the discussions of particular physical
examples has a rather complex structure which will be presented in the
extended text~\cite{lat}.
\end{references}
\end{document}